\documentclass[twocolumn,showpacs,preprintnumbers,prl,aps,apssymb]{revtex4}

\usepackage{graphicx}
\usepackage{dcolumn}
\usepackage{bm}
\newcommand{\ber}{\begin{eqnarray}}
\newcommand{\eer}{\end{eqnarray}}
\newcommand{\bea}{\begin{equation}}
\newcommand{\eea}{\end{equation}}

\begin{document}


\title{Morphological changes of a superconducting phase in a mixed state with a normal current: a multiple scale analysis}

\author{A. Bhattacharyay}
\affiliation{%
Department of Chemistry, University of Warwick, UK\\
}%

\date{\today}

\begin{abstract}
The mixed state of superconducting (SC) and normal (N) phases in one dimensional systems are characterized by several phase slips and localization of the order parameter of the SC phase. The phenomenon is explained on the basis of a complex Ginzburg Landau (CGL) model. We present a simple analysis of the system on multiple scales to capture localization and phase slips when phases coexist.          
\end{abstract}

\pacs{}
\maketitle
\par
Below the critical temperature a one dimensional superconducting channel is driven from superconducting to normal phase by increasing the current passing through it. The phase transition is current driven and the role of excitations to drive such a transition has been explained on the basis of a (CGL) model by Langer and Ambegaokar which was later 
improved by McCumber and Halperin (LAMH) \cite{lan,mac}. Although LAMH approach was phenomenological, similar models have been derived on a microscopic basis \cite{kop}. The LAMH theory is reckoned to have adequately captured the essential thermodynamic features of this current driven transition. However, this classic problem is often revisited from different view points. A recent work that characterizes the periodic appearance of phase slip center (PSC) in the superconducting phase by collision of real eigenvalues at a Hopf bifurcation as a consequence of the {\it PT} (Parity and Time reversal) symmetry of the system is quite interesting \cite{rub}. A closed expression for the temperature and current dependent free energy barrier of clean superconducting micro channels has also been derived recently \cite{zha}.

\par
In the present paper we will investigate the effect of a small normal (Ohmic) current on a superconducting phase in a region where the phases can co-exist. We will employ multiple scale perturbation to find the form of slowly varying amplitude of the SC order parameter. Although the analysis is extremely simple yet reveals an exact form of localization of the SC phase and captures the periodic phase slips. We can also draw a few important conclusions from this very simple analysis. (1) The nonlinearity of the CGL does not play any role whatsoever on the generation of the localization and the dynamic phase slip centers (PSC). (2) Phase slips are a consequence of the spatial variation of the amplitude of the order parameter than a direct consequence of the existence of the Ohmic current. (3) The localization of the SC order is a direct consequence of the existence of the Ohmic current. (4) There are indications that the local nature of the phase slips are a consequence of {\it PT} symmetry of the system in the absence of which it either have not occurred or have been smeared over space. 

\par The CGL model for 1D Superconductor reads as
\begin{eqnarray}\nonumber
&u&(\psi_t+i\mu\psi)=\psi_{xx}+(1-|\psi |^2)\psi,\\
&j&=Im(\psi^*\psi_x)-\mu_x 
\end{eqnarray} 
where $\psi$ is the superconducting order parameter. The system being 1D, x is the distance along the wire, $\mu $ is the electrochemical potential which characterizes the N phase of the system. The $j$ is the current and parameter $u$ is the ratio of order parameter relaxation time to the current relaxation time. The typical value of $u$ is 12 in the strong depairing limit whereas 5.79 for weak depairing \cite{scm}. We will treat $u$ as a constant here. The suffix $t$ and $x$ of $\psi $ indicates corresponding derivatives of it.

\par
Eq.1 has two stationary solutions. (1) $\psi \equiv 0 $, $\mu = -xj$ (for constant $j$) which is the normal state and (2) $\psi=Aexp(iqx)$, $q^2=1-A^2$, $j=A^2q$ which is the superconducting state when $\mu\equiv 0$ \cite{kra}. There is a critical current $j_c$ above which the SC state gives up to the normal state and just below $j_c$ over some range on the current scale the SC phase is metastable and often seen to coexist with the normal state. An extensive numerical analysis by Kramer et al in this region showed localized SC phase \cite{kra} whose profile very much resembles a Gaussian one. The corresponding profile of the $\mu $ is like a straight line passing through the center where the SC profile peaks. In what follows, we will analyze the system in such a region below the $j_c$ (say) where two such phases can coexist.

\par
Since the dominant phase below $j_c$ is the SC phase, let us consider its order parameter $\psi_{sc}$ to have the form $\psi_{sc}=A(X,\tau)[\psi^0+\epsilon\psi^1]$ where the amplitude of the order parameter $A(X,\tau)$ is a function of slow space and time scales $X$ and $\tau$ respectively. We also take into account an additive small variation ($O(\epsilon)$) on the order parameter in the form of $\psi^1$ which supposedly appears in the presence of an equally small ($O(\epsilon)$) presence of a normal current ($\epsilon j_N$). We are interested in getting a form of $A(X,\tau)$ in such a situation. Since we perturb the steady SC state, the time scale under effective consideration here is the slow one only $t = \epsilon \tau$, but we should take multiple length scales as $x=x+\epsilon X$ where the amplitude is the only slowly varying function. 
\par
 In the present analysis we will neglect the role of nonlinearity to enable us apply multiple scale perturbation technique as it is normally done near a linear instability threshold to derive slow amplitude equations.
At this stage, let us modify the model Eq.1 in such a way that does not affect the basic steady state results of the system. We will from now on consider $A^2$ in the place of $|\psi|^2$ in the model where $A$ is exactly the constant amplitude of the steady SC order parameter. This will render the model linear without changing the steady state solutions. So, form now on in what follows we will treat $A$ as a constant if not explicitly mentioned to be function of slow variables.

\par Now, at $O(1)$ we get
\begin{equation}
\psi^0_{xx}+(1-A^2)\psi^0 = 0
\end{equation}
whose solution is the steady superconducting state $\psi_{sc} = A(X,\tau)exp(iqx)$ where at smaller scales $A(X,\tau)$ is effectively constant $A$. Thus, the above made modification of CGL can also be viewed as considering dynamics on widely varying scales.
At $O(\epsilon)$ we get
\begin{eqnarray}\nonumber
\psi^1_{xx}+(1-A^2)\psi^1 &=& [u\frac{\partial A(X,\tau)}{\partial \tau} - i2q\frac{\partial A(X,\tau)}{\partial X}\\ &-& ij_NxA(X,\tau)]exp(iqx),
\end{eqnarray}
where on the left hand side we have absorbed the factor $A$ of $\psi^1$ within $\psi^1$ since $A$ is effectively treated as constate on small scales. The term on the right hand side of the above equation is a secular term. So, it has to vanish for $\psi^1$ to exist since the linear operator (considering $A^2$ in the place of $|\psi^0|^2$) on the left hand side has a zero eigenvalue solution. Thus, we get the linear amplitude equation
\begin{equation}
\frac{\partial A}{\partial t} -\frac{i2q}{u}\frac{\partial A}{\partial x} - \frac{i\epsilon j_N}{u}xA =0
\end{equation}

\par
The $\epsilon$ in the last term appears as a consequence of scale change to the normal ones. Consider the transformation $A=Ae^{-mx^2}$. So, $\frac{\partial A}{\partial x} = -2mxAe^{-mx^2} + e^{-mx^2}\frac{\partial A}{\partial x}$. Putting this in the above amplitude equation we get a selection for $m$ as $m = \frac{\epsilon j_N}{4q}$ which will cancel the terms with a factor $x$ in the amplitude equation. The inverse characteristic length $m$ is positive for positive $\frac{j_N}{q}$ and causes a Gaussian localization of the SC order parameter. Its important to note that for constant normal current $j_N$ the electrochemical potential $\mu$ is a straight line passing through the origin in space and our analytic expression very closely resembles what has been numerically obtained by Kramer et al in \cite{kra}. The localization of the SC phase results as a consequence of the existence of nonzero normal current. An interesting point to note is that the smaller the normal current the wider is the SC phase. The SC state also spreads for larger $q$.  

\par
Let us concentrate on the rest of the amplitude equation. It reads
\begin{equation}
\frac{\partial A(x,t)}{\partial t} = \frac{i2q}{u}\frac{\partial A(x,t)}{\partial x}.
\end{equation}
Consider the form of $A(x,t)$ as $A(x,t)=a(t)exp(\pm \lambda x)$ with $\lambda$ real. So, we readily get $A(t) = exp(i\frac{2q\lambda}{u}t)$ where the amplitude oscillates with a frequency $\omega = \frac{\pm 2q\lambda}{u}$. Since, the amplitude equation is linear, a linear superposition of the solutions are always possible and a general form of the amplitude can be written as
\begin{equation}
A(x,t)=[e^{\lambda x}e^{i\omega t}+e^{-\lambda x}e^{-i\omega t}]e^{-mx^2}
\end{equation} 

\par
This amplitude has a real temporal oscillation at $x=0$ in the form of $Cos(\omega t)$ due to superposition. Since, the amplitude of the two temporally oscillating terms vary every where except at the origin in space we will only get localized real oscillatory solutions at this point. This localized oscillation will cause periodic phase slip at origin when the amplitude vanishes. Interestingly enough, the period of oscillation is not dependent on the $j_N$ and thus phase slips are always there even if we do not consider the presence of the normal current from the very beginning. It only depends on the exponential variation of the order parameter amplitude in space. However, in the presence of the normal current the localization of the amplitude will suppress the exponential divergence of the other part at large distances and thus would make the solution an acceptable one. Nevertheless, one can easily guess that exponential amplitude profiles formed locally due to inhomogeneities or other causes can also cause periodic phase slips.

\par
Another important point to note is the joint variation of sign in the space and time part of the linear solution is actually enabling us to have the localized oscillation. If we had $\omega $ of both the signs corresponding to the same sign of $\lambda $ the superposition would have resulted in a global oscillation. This fact indicates that the symmetry of the system under joint inversion of space and time is important for generation of localized phase slips which has been investigated numerically at length in \cite{rub}. The other very important point revealed by our analysis is that there is no need of the nonlinearity for the generation of the localization and phase slip centers. The nonlinearity definitely plays its role on the stability of these structures as has been revealed by previous works \cite{lan,mac,kop}.

\par
We will conclude by saying that a very simple multiple scale analysis of the of the CGL model for 1D superconductors analytically captures most important characteristics of the system near a region close to the transition from SC to normal phase. (1) It reveals the nonlinearity is not at all responsible for intrinsic generation of localization and PSCs of the system. (2) The localization of the SC phase is a direct consequence of the presence of the normal current whereas the PSCs are not. (3) PSCs will appear even at a zero normal current but with suitable local variation of SC amplitude in space (if the normal current does not indirectly results from the amplitude variation) caused by inhomogeneities or other causes. (4) Symmetry under joint space time inversion seems to play an important role in the localization of the periodic phase slips.

\end{document}